\documentclass[aps,prd,showpacs,amsmath,nofootinbib,superscriptaddress,twocolumn,floats,preprintnumbers]{revtex4}
\usepackage[dvips]{color,graphicx,epsfig}
\usepackage{amsfonts,amssymb,theorem,mathrsfs,times}

\usepackage{natbib}
\usepackage{graphics}
\usepackage{multirow}
\usepackage{amsbsy}
\usepackage{mathrsfs}
\usepackage{footmisc}
\usepackage{hyperref}
\hypersetup{
  colorlinks = true,
  urlcolor = black,
  linkcolor=black,
  citecolor=black
}

\usepackage[dvips]{color,graphicx,epsfig}
\usepackage{amsfonts,amssymb,theorem,mathrsfs,times}
\definecolor{red  }{rgb}{1,0,0}
\definecolor{blue }{rgb}{0,0,1}
\definecolor{green}{rgb}{0,1,0}

\newcommand{\blue}[1]{  {{\textcolor{blue}{  #1}} }  }

\begin{document}
\newcommand{\M}{\mathbb{M}}
\newcommand{\R}{\mathbb{R}}
\newcommand{\HI}{\mathbb{H}}
\newcommand{\C}{\mathbb{C}}
\newcommand{\TI}{\mathbb{T}}
\newcommand{\E}{\mathbb{E}}
\def\etal{{\it et al}.} \def\e{{\rm e}} \def\de{\delta}
\def\dd{{\rm d}} \def\ds{\dd s} \def\ep{\epsilon} \def\de{\delta}
\def\goesas{\mathop{\sim}\limits} \def\al{\alpha} \def\vph{\varphi}
\def\Z#1{_{\lower2pt\hbox{$\scriptstyle#1$}}}

\def\X#1{_{\lower2pt\hbox{$\scriptstyle#1$}}}

\newcommand{\be}{\begin{equation}}
\newcommand{\ee}{\end{equation}}
\newcommand{\bea}{\begin{eqnarray}}
\newcommand{\eea}{\end{eqnarray}}
\newcommand{\nn}{\nonumber}

\def\IR{{\hbox{{\rm I}\kern-.2em\hbox{\rm R}}}}


\renewcommand{\thefootnote}{\fnsymbol{footnote}}
\long\def\@makefntext#1{\parindent 0cm\noindent \hbox to
1em{\hss$^{\@thefnmark}$}#1}
\def\Z#1{_{\lower2pt\hbox{$\scriptstyle#1$}}}


\title{Gauss-Bonnet assisted braneworld inflation in light of BICEP2 and Planck data}

\preprint{CERN-PH-TH-2014-168}

\author{Ishwaree P. Neupane}
 \affiliation{Theory Division,
CERN, CH-1211 Geneva 23, Switzerland}

\begin{abstract}

Motivated by the idea that quantum gravity corrections usually
suppress the power of the scalar primordial spectrum (E-mode) more
than the power of the tensor primordial spectrum (B-mode), in this
paper we construct a concrete gravitational theory in
five-dimensions for which $V(\phi)\propto \phi^n$-type inflation
($n\ge 1$) generates an appropriate tensor-to-scalar ratio that
may be compatible with the BICEP2 and Planck data together. The
true nature of gravity is five-dimensional and described by the
action $S = \int d^5{x} \sqrt{|g|} M^3 \left(- 6\lambda M^2 + R +
\alpha M^{-2} {\cal R}^2\right)$ where $M$ is the five-dimensional
Planck mass and ${\cal R}^2=R^2-4 R_{ab} R^{ab} + R_{abcd}
R^{abcd}$ is the Gauss-Bonnet (GB) term. The five-dimensional
``bulk" spacetime is anti-de Sitter ($\lambda<0$) for which
inflation ends naturally. The effects of ${\cal R}^2$ term on the
magnitudes of scalar and tensor fluctuations and spectral indices
are shown to be important at the energy scale of inflation. For
GB-assisted $m^2\phi^2$-inflation, inflationary constraints from
BICEP2 and Planck, such as, $n_s\simeq 0.9603~(\pm 0.0073)$,
$r=0.16~(+0.06-0.05)$ and $V_*^{1/4} \gtrsim 1.5\times
10^{16}~GeV$ are all satisfied for $ (-\lambda \alpha) \simeq
(3-300)\times 10^{-5}$.


\end{abstract}


\maketitle


\section{Introduction} The 2013 Planck data constrains the scalar
spectral index to $n_s=0.9603\pm 0.0073$~\cite{Planck2013} -- a
result compatible with that predicted by primordial cosmic
inflation~\cite{Guth1980}. The BICEP2 (Background Imaging of
Cosmic Extragalactic Polarization 2) collaboration~\cite{BICEP2}
reports on the detection of inflationary gravitational waves
(B-mode polarization), which confirms yet another prediction of
inflation. Such results from Planck and BICEP2 offer a rare
opportunity to directly test theoretical models, including
inflation. BICEP2 results in particular indicate toward some new
physics around the energy scale of inflation, $\rho^{1/4} \sim
1.5\times 10^{16}~{\rm GeV}$, which is in the order of symmetry
breaking scale of the grand unified theory. The tensor
fluctuations in the cosmic microwave background (CMB) temperatures
at large angular scales are larger than those predicted for
inflationary models based on Einstein gravity. Specifically, the
ratio of tensor-to-scalar perturbations reported by BICEP2
collaboration, $r=0.19^{+0.07}_{-0.05}$ (or
$r=0.16^{+0.06}_{-0.05}$ after subtracting an estimated
foreground), is larger than the bounds $r < 0.13$ and $r < 0.11$
reported by WMAP~\cite{Komatsu:2010} and Planck~\cite{Planck2013}.
Some part of this discrepancy may be accounted for by postulating
that the value of $r$ measured by BICEP2 at $\ell\simeq 60$
corresponds to a smaller field of view of the sky where an
inflationary gravitational waves signal would be expected to peak,
whereas the value of $r$ measured by Planck at $\ell\simeq 30$
corresponds to a larger field of view of the sky where $r$ gets
attenuated. In this paper we identify a concrete gravitational
theory in which the value of $r$ gets enhanced due to quantum
gravity corrections or higher-curvature terms. The latter suppress
the scalar primordial power spectrum (PPS) more than the tensor
PPS at high energies (compared to the results in general
relativity) for $\phi^n$-type potentials ($n \ge 2/3$). This is
one of the key results of this paper.

One way to accommodate quantum effects of gravity is to include
curvature-squared terms in a gravitational action. Such terms
arise from the low energy effective action of string theory and/or
as the $1/{\cal N}$ corrections in the large ${\cal N}$ limit of
some gauge theories~\cite{Aharony:1999}. The Gauss-Bonnet (GB)
combination of curvature invariants, ${\cal R}^2 = R^2 - 4 R_{ab}
R^{ab}+ R_{abcd} R^{abcd}$, is of particular relevance in five
dimensions, since it represents the unique combination that leads
to second-order gravitational field equations and hence to
ghost-free solutions in flat as well as in curved
spacetimes~\cite{CNW-01}.

Inflationary constraints from Planck data have been based on
general relativity. An exponential potential (also called power
law inflation) is {\it not} favored by Planck data for two
reasons; one reason is that the value of $r$ is relatively large,
$r \simeq -8(n_s-1)\simeq 0.32$, which is beyond $2\sigma$
confidence level of the Planck data ($r<0.26$, 95\% confidence).
The second reason, though not limited to exponential potential, is
that inflation would not end without an additional mechanism to
stop it. A question of great significance is: What happens if the
early universe is better described by the following
Einstein-Gauss-Bonnet gravitational action in five dimensions?
\bea S &=& \int_{\cal M} d^5{x} \sqrt{|g|} M^3 \left(- 6 \lambda
M^2+
R + {\alpha\over M^{2}} {\cal R}^2\right) \nn \\
&{}& \quad + \int_{\partial {\cal M}} d^4{x} \sqrt{|\tilde{g}|}
\left({\cal L}_m +{\cal L}_\phi - \sigma\right),\eea where $M$ is
the five-dimensional Planck mass, ${\cal L}_m$ (${\cal L}_\phi$)
is matter (scalar) Lagrangian and $\sigma$ is the brane tension
(or a cosmological constant in four dimensions). The above action
is consistent with {\it braneworld} realisation~\cite{RS2} of
string and M theory, according to which all elementary particles,
gauge fields and fundamental scalars live within a
four-dimensional ($3$ dimensions of space and $1$ dimension of
time) membrane, or ``brane,'' while the effect of gravity extends
along the fifth dimension. The condition $-\lambda \alpha < {\cal
O} (1/10)$~\cite{CNW-01} is required for the stability of
classical solutions under perturbations, which also guarantees a
suppression of higher powers of curvature tensors. Inflationary
constraints provide a more stringent bound, $(-\lambda\alpha) <
{\cal O}(10^{-3})$. In this paper, we show that the above
mentioned model leads to amazingly simple four-dimensional
universe that resembles in many ways the one observed by BICEP2
and Planck.

\section{Modified Friedmann equations}

We are interested in cosmological solutions, so we write the 5D
metric {\it ansatz} as
\be \label{5D-brane} ds^2 = - N(t,y)^2 \,dt^2+ A^2(t,y)
d\Omega_{3,\kappa}^2 + B(t,y)^2 \, dy^2. \ee The use of gauge
$N(t,y=0)=N\Z{0}\equiv 1$ implies $\dot{A}= a H N$ (where $a\equiv
A\Z{0}$ is the scale factor of the
Friedmann-Lama\^itre-Robertson-Walker universe) and $t$ is the 4D
proper time. The set of four-dimensional field equations are given
by
\begin{eqnarray}
 X\left[ 1 + {4\alpha H^2\over M^2} \left( 1
  - {X^2\over 3 H^2}+{k\over a^2 H^2} \right)\right] &=& - {\left(\rho+\sigma\right)\over
  6M^3},~~
  \label{junction1}\\
 X \left[1+{4\alpha H^2\over M^2} \left(1+{\dot{H}\over H^2}-{X
Y\over 3H^2}\right)\right]
\nn \\
 + {Y\over 2}\left[1+{4\alpha H^2\over M^2} \left(1-{X^2\over
3H^2}+{k\over a^2 H^2}\right)\right] &=&
{\left(p-\sigma\right)\over 4 M^3},\label{junction2}
\end{eqnarray} (see also Refs.~\cite{Davis:2002,Maeda:2003}) where %
\be X \equiv {A'\vert_{y=0}\over {a B\Z{0}}} = -\sqrt{H^2 + \psi^2
M^2+{k\over a^2}}, \quad Y\equiv {N'\vert_{y=0}\over B\Z{0}}.
\nn\ee 
\be \psi^2= {1-\sqrt{\Delta}\over 4\alpha} \quad {\rm and}\quad
\Delta\equiv 1+8\lambda \alpha + {8\alpha\over a^4} {{\cal E}\over
M^2}.\nn\ee ${\cal E}$ is a measure of bulk radiation energy,
which is proportional to the mass of a 5D black hole. $\psi$ is a
dimensionless measure of bulk curvature. In the case the 5D
spacetime is anti-de Sitter ($\lambda<0$) and the GB coupling
constant is positive ($\alpha>0$), such that $\Delta< 1$, the
${\cal R}^2$-type corrections would lead to graceful exit from
inflation for a number of scalar potentials.



To study inflation, we will ignore the term $\kappa/a^2 H^2$,
which is justified from the viewpoint that inflation would stretch
any initial curvature of the universe to near
flatness~\footnote{The universe may be slightly open at present,
$(-\Omega_\kappa) \equiv {\kappa/a^2 H^2}= (10^{-2}\sim 10^{-3})$,
if so, $\kappa/a^2 H^2$ becomes important at low energies.}. The
Friedmann equation (\ref{junction1}) can be written as \be H^2=
{M^2 \psi^2\over
\beta}\left[(1-\beta) \cosh\varphi -1\right],\label{main-Fried1}\ee 
\be \varphi \equiv {2\over 3} \sinh^{-1}\left( \sqrt{\alpha\over
2} {\rho+\sigma \over M^4} {1\over
\Delta^{3/4}}\right),\label{def-varphi}\nn \ee where $\beta\equiv
4\alpha \psi^2=1-\sqrt{\Delta}$. In the early universe, most
energy is present in the $\phi$-field, which means \be \rho\simeq
\rho_\phi = {4(1-\beta)^{3/2}\over (2\beta)^{1/2}} \psi M^4
\sinh(3\varphi/2)- \sigma. \ee The brane tension $\sigma$ is {\it
not} fine-tuned except in the Randall-Sundrum limit~\cite{RS2}
$\rho \to 0$ (vacuum dominated universe) for which $\sigma = 2\psi
M^4 (3-\beta)\delta$ and $\delta= 1$~\cite{Ish02a}. The natural
choice is $0\le \delta \le 1$. The scalar-matter density
$\rho_\phi={1\over 2}\dot{\phi}^2 + V(\phi)$.

\section{GB assisted inflation}

The inflaton equation of motion is \be \ddot{\phi}(t)+ 3 H(t)
\dot{\phi}(t) + V_\phi=0, \ee where $V_\phi\equiv dV/d\phi$.
Scalar fluctuations generated from inflation include two types of
contributions, one coming from quantum fluctuations of the
$\phi$-field and the other from massive Kaluza-Klein (KK) modes
from the bulk. The massive scalar modes with mass $m\Z{\rm KK}>
3H/2$, which are too heavy to be exited during a slow-roll
inflationary era, are rapidly oscillating and their amplitudes are
strongly suppressed on largest scales~\cite{Maartens:1999}. If the
energy density of $\phi$-field is very large during inflation,
such as $\rho^{1/4}\gtrsim 1.5\times 10^{16}~{\rm GeV}$ (as
inferred by Planck and BICEP2 data), then most contributions to
scalar perturbations would arise from the $\phi$-field. Moreover,
after a few e-folds of inflation the effect of bulk radiation
energy becomes negligibly small in which case $\Delta$ ($\beta$)
is constant. Under these two approximations, which are reasonably
good, the amplitude of scalar (density) perturbations is given by
\be A\Z{S}^2\equiv {4\over 25} {\cal P}_{\rm sca}(k) ={9\over
25\pi^2} {H^6\over V_\phi^2}.\ee The amplitude of primordial
tensor fluctuations depends on the energy scale of inflation and
nature of quantum gravitational fields that generared gravitons --
the elemenatry particles that mediate the force of gravity. The
normalized amplitude of primordial tensor perturbations was
previously obtained in Refs.~\cite{Dufaux:2004,Bouhmadi-Lopez},
which in our notation ($\mu\equiv \psi M$), reads as \be A\Z{T}^2
\equiv {1\over 25} {\cal P}_{\rm ten}(k)= {2\over 25} {\psi\over
M^2 {\cal A}} \left({H\over 2\pi}\right)^2 ,\ee \be {\cal A}\equiv
(1+\beta)\sqrt{1+x^2} - (1-\beta) x^2 \sinh^{-1} {1\over x},\nn
\ee where $x\equiv H/(\psi M)=\beta^{-1/2} \left[(1-\beta)
\cosh\varphi-1\right]^{1/2}$ is a dimensionless measure of Hubble
expansion rate. The five-dimensional impact on the scalar and
tensor power spectra is largely characterised through a
modification of Hubble expansion rate as given in
Eq.~\ref{main-Fried1}. The power of the scalar and tensor
primordial spectra can be calculated approximately in the
framework of the slow-roll approximation by evaluating the above
equations at the value $\varphi=\varphi_*$ where the mode $k_*=a_*
H_*$ crosses the Hubble radius for the first time. On the usual
assumption that $H$ is nearly constant throughout inflation, the
amplitude of scalar density perturbations has some scale
dependence due to a small variation in $V_\phi$, while the tensor
perturbations are roughly scale independent.

The number of e-folds of inflation $N\equiv \int H dt$ is given by
\bea N \equiv \int_{\varphi_*}^{\varphi_e} H {dt\over d\phi}
{d\phi\over dV} {d V\over d\varphi} d\varphi \simeq  3
\int_{\varphi_e}^{\varphi_*} {H^2\over V_\phi^2} \left(dV\over
d\varphi\right) d\varphi, \nn \eea where the equality holds in the
slow-roll approximation $\ddot{\phi} \ll 3 H(t) \dot{\phi}$ and
subscript ``e" refers to the end of inflation. The simplest class
of inflationary models is characterized by monomial potentials of
the form $V(\phi)= m^{4-n} \phi^n$. For slow-roll inflation, the
first two slow-roll parameters $\epsilon \equiv - \dot{H}/H^2$ and
$\eta = V_{\phi\phi}/(3H^2)$ are evaluated to be \bea \epsilon &=&
{(2+n)(1-\beta)\,I(\varphi) \over 2\big[N(2+n) +
n\big]}\nn \\
&{}& \times {\sinh\varphi \left[\sinh(3\varphi/
2)-c\right]^{2-2/n}\over
\left[(1-\beta)\cosh\varphi -1\right]^2 \cosh(3\varphi/2)},\\
\eta &=& {3(n-1)(n+2) I(\varphi)\over 2 n\big[N(2+n)+n\big]}
{\left[\sinh(3\varphi/2)-c\right]^{1-2/n}\over
(1-\beta)\cosh\varphi-1},\nn \\ \eea where $c\equiv
(3-\beta)\beta^{1/2}(1-\beta)^{-3/2}\delta/\sqrt{2}$. Here we give
the explicit expression of $I(\varphi)$ for $n=2$, $n=1$ and
$n=2/3$:
\bea I(\varphi) &=& \varphi-{2\beta\over 3}
\ln\left(e^\varphi-1\right) +(1-\beta) (\cosh \varphi-1) \nn \\
&{}& \quad + {3-\beta\over 3}\big[\ln 3-
\ln\left(e^{2\varphi}+e^\varphi+1\right)\big], \nn\\
I(\varphi) &=& {(1-\beta)\over 5}\sinh{5\varphi\over 2}-{2\over
3}\sinh{3\varphi\over 2} +(1-\beta)\sinh{\varphi\over 2}, \nn \\
I(\varphi) &=& {1-\beta\over
48}\left(3\cosh4\varphi+6\cosh2\varphi-8\cosh3\varphi-1\right).\nn
\eea
The scalar spectral index is given by \be n_s-1 \equiv {d\ln
A_S^2\over d\ln k}\Big|_{k =aH}= -6\epsilon + 2\eta.\ee The
tensor-to-scalar ratio $r\equiv 4 {\cal P}_{\rm ten}/{\cal P}_{\rm
sca}$ is given by \bea r &=&{8(2+n)\,I(\varphi) \over N(2+n) + n}
{(1-\beta)^{3/2} |2\beta|^{1/2}\over {\cal A}} \nn
\\
&{}& \quad \times {\left[\sinh(3\varphi/ 2)-c\right]^{2-2/n}\over
\left[(1-\beta)\cosh\varphi-1\right]^2}.\eea The above results are
valid for $0< |\beta|\ll 1$ and $V(\phi) \gg \sigma$ and they
improve the expressions given in~\cite{Sami04b} where the limit
$\beta\to 0$ was taken but erratically. The results corresponding
to an exponential inflation $V(\phi)\propto
\exp[\gamma\phi/M\Z{P}]$~\cite{Lidsey:2003} are obtained by taking
$n\to \infty$. Inflation begins at $\varphi=\varphi_*$ and
depending on the form of the potentials we observe that
$\varphi_*\sim (0.6-2.5)$; inflation begins at a larger field
value for an exponential potential. As shown in Fig.~1, inflation
has a natural exit ($\epsilon>1$) only if $\beta>0$ (or
$\lambda<0$ for $\alpha>0$).

\begin{figure}[!ht]
\centerline{\includegraphics[width=3.0in,height=1.8in]
{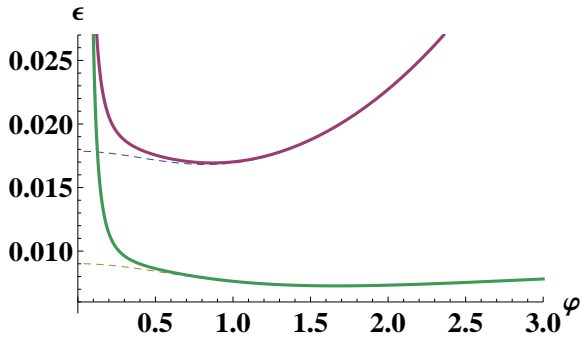}} \caption{$\epsilon$ vs $\varphi$:
$V\propto \exp[\gamma\phi/m\Z{P}]$ (upper plot) and $V\propto
\phi^2$ (lower plot) with $N_*=55$, $\beta=0$ (dotted) and
$\beta=10^{-3}$ (solid) lines.}\label{epsilon}
\end{figure}

\begin{figure}[!ht]
\centerline{\includegraphics[width=3.5in,height=2.0in]
{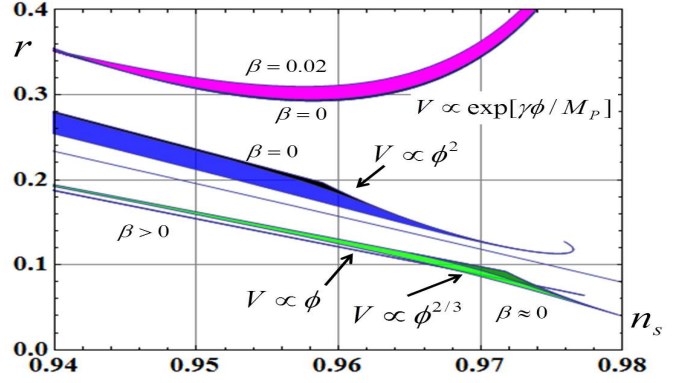}} \caption{The GB-assisted
$\phi^n$-inflation: tensor-to-scalar ratio vs scalar spectral
index with $N_*=60$, $0< \beta \le 0.02$ and $\delta = 0$.}
\end{figure}

\begin{figure}[!ht]
\centerline{\includegraphics[width=3.5in,height=2.0in]
{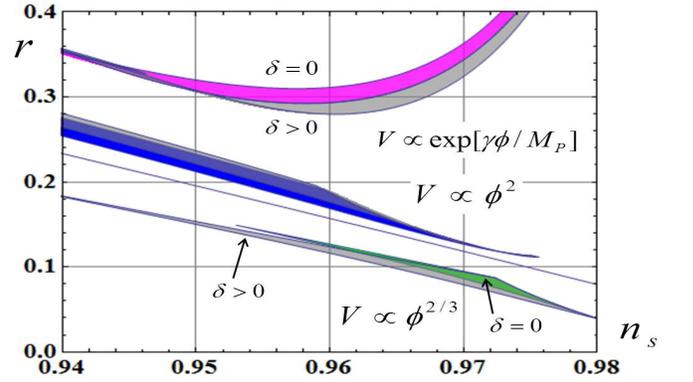}} \caption{As in Fig.~2 but $0 \le \delta
< 0.5$.}
\end{figure}

In Fig.~2 color bands correspond to $\beta>0$ and the bold lines
correspond to $\beta\approx 0$. The effect of $\beta$ on monomial
potentials $V\propto \phi^n$ is {\it not} uniform for all values
of $n$. Especially, for $m^2\phi^2$ inflation, a positive $\beta$
suppresses the amplitude of primordial tensor (gravitational
waves) fluctuations more than the scalar (density) primordial
fluctuations. This effect is opposite for other values of $n$,
including $n= 1$ and $n=2/3$. In all cases the value of $r$ is
greater than their values in general relativity
(GR)~\cite{Planck2013}. Note that the limit $\beta\to 0$
corresponds to Randall-Sundrum braneworld
cosmology~\cite{Maartens:1999}, {\it not} to the GR limit. For
$m^2\phi^2$ inflation in GR, $n_s\sim 0.96$ corresponds to $r\sim
0.157$ (see e.g.~\cite{Creminelli2014} or the single solid line in
Figs.~2 and 3). For $\beta\approx 0$, $n_s=0.96$ implies
$r=0.1742$ ($0.1885$) for $N_*=50$ ($60$). The value of $r$
decreases once $\beta$ is increased; for example, for $\beta\simeq
10^{-4}$, $n_s\simeq 0.96$ implies $r\simeq 0.1741$ ($0.1822$) for
$N_*=50$ ($60$). These values are close to the central value of
$r$ reported by BICEP2 collaboration. For $V\propto
\phi^{2/3}$-inflation, the value of $n_s$ ($r$) is relatively
large (small). For $\beta\approx 0$, $n_s\sim 0.971-0.980$ and
$r\sim 0.04-0.1$, which both are away from the mean values of
$n_s$ and $r$ reported by Planck and BICEP2 collaborations. If
$\beta>0$, as shown in Fig.~2, a smaller (larger) value of $n_s$
($r$) can also be obtained; viz, $(n_s, r)= (0.96, 0.13)$. This
kind of suppression in scalar power with a larger tensor-to-scalar
ratio at higher energies (in Gauss-Bonnet regime) can help to
reconcile the Planck and BICEP2 data in a single framework.

For a small coupling constant like $\beta< 0.001$, we get $c<
0.067$; the effect of brane tension is small if the energy scale
of inflation is large. In Fig.~3, color bands correspond to
$\delta=0$ and the grey bands to $\delta=1/2$. A positive $\delta$
lowers the value of $r$ for an exponential potential, while it
increases $r$ for $m^2\phi^2$ inflation. For $V(\phi)\propto
\phi$, the values of $r$ and $n_s$ do not depend on $\sigma$ (or
$\delta$) and they depend on $N_*$ modestly; for $n_s\simeq 0.96$,
$r \simeq 0.119 - 0.121$ with $N_* \sim 50-60$.

\begin{figure}[!ht]
\centerline{\includegraphics[width=3.5in,height=2.0in]
{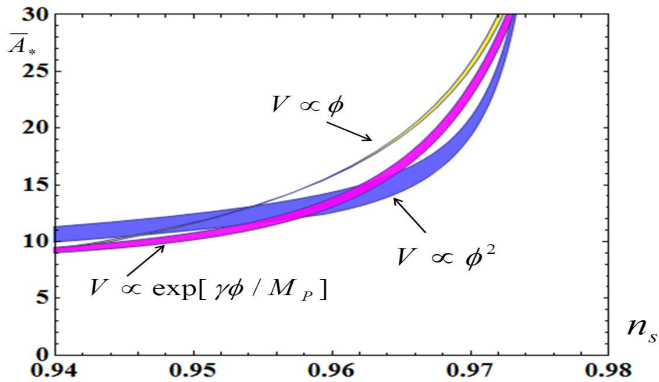}} \caption{The COBE normalized amplitude of scalar
perturbations $\bar{A}_*\equiv (M\Z{P}/M)^6 \times A_*$ vs $n_s$
with $N_*=60$ and $0.02 > \beta>0$.}\label{As-vs-ns}
\end{figure}

In order to constrain the model parameters we use the COBE
normalization for amplitude of scalar perturbations used by Planck
collaboration, $A_*\simeq V^3/(12\pi^2 M\Z{P}^6 V_\phi^2)\simeq
22\times 10^{-10}$. As shown in Fig.~4, when $n_s\simeq 0.96 $, we
get
$$ A_* \simeq 15 \times \left(M/M\Z{P}\right)^6 \quad \rightarrow \quad
M\simeq 5.51\times 10^{16}~{\rm GeV}.$$ Further, $\beta$ and
$\psi$ may be constrained by using the dimensional reduction
relation $ \psi m\Z{P}^2 = (1+\beta) M^2$~\cite{CNW-01} between
the four- and five-dimensional Planck masses. For GB assisted
$m^2\phi^2$ inflation, with $N_*\simeq 55$ and $n_s\simeq 0.96$,
we observe that \be 12 < \xi^2< 48, \quad \xi^2 \equiv {4\times
10^4\cdot \alpha\cdot m^2\over M^2}. \ee A deviation from
$n_s\simeq 0.96$ changes this bound slightly; for example, if
$n_s\simeq 0.963$, then $25 <\xi^2<60$. During inflation $m\gtrsim
H \simeq 1\times 10^{14}~{\rm GeV}$, which means $\alpha\sim
91-364$ and $\beta \sim (1-4)\times 10^{-4}$. For a small-mass
inflaton field, $m\simeq 2\times 10^{13}~{\rm GeV}$, the bound on
$\beta$ is tighter, $\beta \sim (25-130)\times 10^{-4}$. These
numbers are compatible with current observations related to
inflationary era for a wide range of the energy scale of inflation
and the number of e-folds. During the early phase of inflation,
$V=m^2 \phi_*^2 \simeq (1.5\times 10^{16}~{\rm GeV})^4$, which
implies that $\phi_*\simeq 0.89\times M\Z{P}$; we have a stage of
inflation at $\phi_* \lesssim M\Z{P}$. Inflation ends at
$\varphi=\varphi_e\simeq 0.01$, or when $V_{\rm end}^{1/4}\simeq
0.16\times M = 8.9\times 10^{15}~{\rm GeV}$ and $\phi_{\rm
end}\simeq 0.33~M\Z{P}$. It follows that $\Delta\phi\equiv
\phi_*-\phi_{\rm end}$, the change in $\phi$ after the scale $k_*$
leaves the horizon, $\Delta\phi \simeq 0.56\,M\Z{P}$. This
estimates agrees with Lyth's recent discussion in~\cite{Lyth:2014}
and there is no Super-Planckian excursion of the inflaton field.
In fact, the Lyth bound $\Delta\phi \gtrsim M\Z{P}\sqrt{r/ 4\pi} $
may not apply to Gauss-Bonnet assisted inflation since the value
of $r$ varies with both the energy scale of inflation $\varphi_*$
and the number of e-folds.

\section{Discussion and some comments on the literature}

Though Ref.~\cite{Maeda:2003} obtained the set of 4D equations
using a covariant formalism, their equations were not written in a
form from which one can easily obtain the expression of Hubble
squared parameter (except in the $\alpha\to 0$ limit). There are
no fundamental disagreements with results in
Refs.~\cite{Davis:2002, Maeda:2003}, which are perhaps correct.
The set of equations given in~\cite{Maeda:2003} are in an abstract
form, which are less useful at least for studying inflationary
solutions since the relation between the Hubble-squared parameter
and the scalar-matter density or the dimensionless scale $\varphi$
related to the scale of inflation was not established. We have
presented results in terms of the model parameters like $\psi$,
$\beta$ and $M$ as they can be linked to inflationary variables.

The plots in Ref.~\cite{Sami04b} are not quite right (except the
first one) since $\beta\to 0$ limit was taken in most of their
discussions after Eq. (18). In this limit, one would be studying
Randall-Sundrum type braneworld inflation; in fact, $\beta=0$
solutions do not characterize the full effect of $R^2$ terms,
since the effects of $R^2$ corrections cannot be accommodated just
by letting $\varphi$ run. One more drawback in the analysis of
Ref.~\cite{Sami04b} is that the authors used the RS type tuning
for brane-tension, $\sigma = 2\psi M^4 (3-\beta)$, which only
holds in the limit $\rho_\phi\to 0$, $H\to 0$ and ${\cal E}\to 0$
but not if any of these quantities is not zero. It is important to
realize, in the present model, that inflation ends only when
$\beta$ is positive, but not when $\beta=0$. This is an important
difference from Ref.~\cite{Sami04b}. More importantly, we have got
a success to constrain the model parameters like $M$, $\beta$, and
or $\lambda\alpha$ for the first time by using inflationary
constrains, such as, the COBE normalized amplitude of scalar
fluctuation and spectral indices. This result is truly remarkable.

The major outcomes of this paper are the results given in
Eqs.(10)-(14) and the plots in Figs. 1 to 4. For a completeness,
we also expressed the 4D field equations in a form most
appropriate, which were known before in one or another form.

We have also shown that a steep inflation may be compatible with
the BICEP2 results (within $2\sigma$ CL, $r<0.27$) provided that
the number of e-folds of inflation $N\gtrsim 70$ and the brane
tension is also large. This improves the earlier discussions in
Ref.~\cite{Lidsey:2003}. In this last reference, some constraints
on $n_s$ were derived for an exponential potential for different
values of $N$. Constraints on $r$ that are compatible with
constraints on $n_s$ were not considered there -- a parametric
plot between $n_s$ and $r$ would help us to compare and confront
theoretical results with Planck and BICEP2 constraints for
inflationary parameters as discussed above.

Here we make one more comment. As long as the GB coupling is
nonzero (no matter how tiny), one would never go to RS regime
because the RS regime means $\alpha=0$ absolutely. It is not true
that inflation begins in the GB regime and ends in the RS regime;
the drop in energy scale $V^{1/4}$ during inflation is generically
only an order of magnitude difference, which means inflation can
occur solely during a phase where $H^2 \propto
(\rho+\sigma)^{2/3}$~\cite{Dufaux:2004}. At a later epoch $H^2$
scales as $(\rho+\sigma)^2$ and this scaling relation can be seen
both with $\alpha=0$ and $\alpha\ne 0$. The ${\cal R}^2$-type
corrections (of a Gauss-Bonnet form) would lead to graceful exit
from inflation for a number of scalar potentials, provided that
$\beta>0$. Moreover, these corrections are important at the
earliest epoch, though their contributions diminish rapidly after
inflation (more precisely, after reheating) all the way to the
epochs of baryogenesis, nucleosynthesis and at the present epoch.

\section{Conclusion}

In this paper we have identified a gravitational theory where
inflation has a natural exit. For $V\propto \phi^n$-type inflation
with $n\ge 1$, it is shown that the Gauss-Bonnet term ${\cal R}^2$
can generate a suppression in scalar power at large scales along
with reasonable amplitudes of primordial scalar and tensor
perturbations ($r\sim 0.12 - 0.20$, $n_s\simeq 0.96$). If BICEP2
is going to confirm their reported result that
$r=0.19^{+0.07}_{-0.05}$~\cite{BICEP2} then an exponential
inflation can be compatible with the result only if the brane
tension or the bare cosmological constant is nonzero and the
number of e-folds of the cosmic inflation is significantly large,
$N_*\gtrsim 70$. The GB-assisted $m^2\phi^2$ inflation is in
agreement with BICEP2 for a wide range of the energy scale of
inflation and number of e-folds $47< N_*<65$. The $m^2\phi^2$
inflation fits better with the BICEP2 result as compared to other
forms of scalar potentials, such as, $V\propto \phi^{2/3}$ and
$V\propto \phi$. A GB-asssisted natural inflation model
characterized by the potential $V=V\Z{0} \left(1\pm
\cos(n\phi/M\Z{P})\right)$ which approximates the $m^2\phi^2$
potential for $n\ll 1$ is also compatible with
BICEP2~\cite{Ish2014b}. 

The gravitational theory discussed in this paper offer simplest
and at the same time important mechanism to generate a larger
tensor-to-scalar ratio (as compared to the results in GR) that can
consistently address the apparent discrepancy between the Planck
upper bound and the BICEP2 detection of $r=0.16^{+0.06}_{-0.05}$
(after subtracting an estimated background) because the
GB-assisted inflation leads to a scenario where tensor
perturbations are roughly scale independent, while some scale
dependence of  scalar density perturbations could relax the Planck
constraint and bring the two results into agreement.

The GB coupling constant $\alpha$ is tightly constrained in
combination along with another constant $\lambda$ associated with
the curvature of the five-dimensional spacetime, namely
$(-\lambda\alpha)\simeq (3-300)\times 10^{-5}$, making a
prediction that could be confirmed or falsified by a future
detection of some scale dependence of scalar density perturbations
and hence the measurement of a small but nontrivial running of the
scalar spectral index and possibly a small non-Gaussianity
parameter. This paper makes a major contribution in the field of
inflationary cosmology by limiting coupling parameters of
Einstein-Gauss-Bonnet gravity that are compatible with BICEP2 and
Planck constraints for primordial cosmic inflation.

\blue{The model investigated here is a natural generalization of
RS model (also called 5-dimensional warped geometry theory). In
string theory, like in type II theory, one adds a five sphere
$S^5$ to get a ten-dimensional space. The five-sphere is usually
related to the scalars, fermions and anti-symmetric form fields in
the super-symmetric Yang-Mills theory. For other theories the
sphere is replaced by other manifolds, or it might even not be
there. It is logical to assume that the effect of scalar(s) is
encoded in the four-dimensional effective scalar Lagrangian as
implicitly assumed in the paper.}


\section*{Acknowledgements}

It is a pleasure to thank Ed Copeland, Naresh Dadhich, Radouane
Gannouji, Shinji Mukohyama, M Sami, and Subir Sarkar for
insightful discussions. I am grateful to Oxford Theory and
Astrophysics Groups and Nottingham University Theory Group for
their hospitality during my visits. This work was supported by the
Marsden Fund of the Royal Society of New Zealand.

\end{document}